\documentclass[useAMS,usenatbib]{mn2e}
\usepackage{graphicx}
\usepackage{array,longtable}
\usepackage{txfonts}
\usepackage{natbib}
\usepackage{float}
\usepackage[T1]{fontenc}
\usepackage{aecompl}
\def\actaa{Acta. Astronom.} %
\def\aj{AJ}%
\def\apj{ApJ}%
\def\apjl{ApJ}%
\def\aap{A\&A}%
\def\mnras{MNRAS}%
\def\pasp{PASP}%
\title[On the Correlation Between Metallicity and the X-Shaped Morphology of the Milky Way Bulge]{On the Correlation Between Metallicity and the X-Shaped Morphology of the Milky Way Bulge}
\author[Nataf, Cassisi, \& Athanassoula]{D. M. Nataf$^1$\thanks{Email: david.nataf@anu.edu.au}, S. Cassisi$^{2}$, E. Athanassoula$^{3}$
\vspace*{6pt}\\
$^{1}$Research School of Astronomy and Astrophysics, The Australian National University, Canberra, ACT 2611, Australia \\
$^{2}$INAF-Osservatorio Astronomico di Collurania, Via M. Maggini, I-64100 Teramo, Italy \\ 
$^{3}$Aix Marseille Universite, CNRS, LAM (Laboratoire d'Astrophysique de
Marseille) UMR 7326, 13388, Marseille, France \\ }
\begin{document}
\include{journaldefs}
\date{Accepted ...... Received ...... ; in original form......   }

\pagerange{\pageref{firstpage}--\pageref{lastpage}} \pubyear{2014}
\maketitle
\label{firstpage}

\begin{abstract}
We demonstrate that failure to properly account for stellar evolution can bias results in determinations of the spatial morphology of Galactic bulge stars,  focusing on the question of whether or not the X-shape is more pronounced among the more metal-rich stars than among the metal-poor stars. We argue that this trend, a result recently claimed by three separate groups, may have been overestimated as it is relatively easier to detect a bimodality in the distance distribution function at higher metallicities. This is due to three factors. First, the intrinsic colour of red clump and red giant stars vary with metallicity, at the level $d(V-I)_{RC}/d\rm{[M/H]} \approx 0.25$ mag dex$^{-1}$, and thus the ratio of red clump to  red giant stars within a spectroscopic sample will depend on the photometric selection of any investigation. Second, the duration of ascent of the red giant branch goes down and the red clump lifetime goes up as metallicity increases, which has the effect of increasing the ratio of red clump to red giant stars by as much as $\sim$33\% over the range of the bulge metallicity-distribution function. Finally, over the same metallicity interval, the effective number of red giant branch bump stars is predicted to increase by $\sim$200\%, and their presence becomes degenerate with the observational parameters of the two red clumps, creating an illusory increase in signal-to-noise for a second peak in the distance modulus distribution. 
\end{abstract}
\maketitle


\section{Introduction}
\label{sec:introduction}
The apparent magnitude distribution of red clump (RC) stars toward Galactic bulge sightlines  that are at least $\sim5$ degrees apart from the plane is bimodal \citep{2010ApJ...721L..28N,2010ApJ...724.1491M}, an artefact of an  excess in the orbital distribution of bulge stars that would appear X-shaped if the Galactic bar were viewed side-on \citep{2012ApJ...756...22N,2012ApJ...757L...7L}. Given that the class of orbits that contributes to this morphology, predominantly trapped by the x1 tree of families, is sharply sensitive to the Galactic gravitational potential \citep{2002MNRAS.337..578P,2003MNRAS.341.1179A}, and that the kinematics of bulge stars have been shown to be correlated to metallicity \citep{2008A&A...486..177Z,2010A&A...519A..77B,2013MNRAS.432.2092N}, mapping how the strength and extent of the X-shape correlates with metallicity could constrain formation and evolution models of the Galaxy. For example, it has been suggested that: 
\begin{quotation}
\noindent Stars supporting the X-shape would primarily be disc stars, the latest ones captured into resonance that were in the mid-plane prior to their capture into resonance. As they were all disc stars, they should have similar metallicity and that typical of the disc just exterior to resonance - \citet{2014MNRAS.437.1284Q}. 
 \end{quotation}
 Such a prediction can only be compared to observations that include both metallicity and kinematic information, thus requiring an accounting of the systematics thereof. 

In that regard, we identify four claims linking the X-shaped morphology of high-latitude bulge stars to metallicity. \citet{2012ApJ...756...22N} used a combination of spectroscopic and photometric data along the bulge minor axis to argue that the X-shape is stronger for stars with [Fe/H]$\geq0$ than for stars with $-0.5 \leq$ [Fe/H] $< 0.0$, with the X-shaped morphology disappearing among stars with [Fe/H] $\lesssim -0.50$.  \citet{2012A&A...546A..57U} independently argued, also by means of a combination of spectroscopic and photometric data, that stars with metallicity [M/H] $\leq -0.20$ do not show the split red clump, in contrast to stars with  [M/H] $\geq -0.20$. This demarcation has also been recently argued in separate conference presentations by Manuela Zoccali\footnote{http://www.ctio.noao.edu/noao/conference/Presentations} and Alvaro Rojas-Arriagada\footnote{http://www.sexten-cfa.eu/images/stories/conferenze2014/bulge/talks/Formevogalaclu-program.pdf} to be confirmed by data from the Gaia-ESO survey \citep{2012Msngr.147...25G}. \citet{2013ApJ...776L..19D} have argued that bulge RR Lyrae stars (with a mean metallicity [Fe/H]$\approx-1.0$, see \citealt{2012ApJ...750..169P}) are distributed as a spheroid, and not as a bar, unlike the more metal-rich RC stars.  The combination of these four works suggests an open-and-shut case: The X-shape of the Galactic bulge is most prominent among the most metal-rich stars, progressively becoming weaker with decreasing metallicity, disappearing entirely for stars with [Fe/H] $\lesssim -0.50$. 

In this Paper, our aim is neither to confirm nor refute this claim, but to argue for further diligence in treating systematics that emerge due to the convolution of stellar physics with Galactic dynamics, this convolution being that which is ultimately observable. We use a combination of stellar and dynamical models (discussed in Section \ref{sec:Models}) to show that even if the X-shape were uniformly prominent among stars of all metallicities, it would still appear more prominent with increasing metallicity due to a combination of up to three factors. These are the metallicity-dependence of the colour of the RC (Section \ref{sec:Systematic1}), the increase in the ratio of RC to red giant (RG) stars with increasing metallicity (Section \ref{sec:Systematic2}), and the effect of the red giant branch bump (RGBB) (Section \ref{sec:Systematic3}).

\begin{table*}
\caption{\large Predicted parameters for the combined red giant  branch + red clump + asymptotic giant branch luminosity function calculated from BaSTI\textsuperscript{\ref{foot:1}} isochrones \citep{2004ApJ...612..168P,2007AJ....133..468C}, as a function of the metallicity [M/H], [$\alpha$/Fe], the age $t$ in Gyr, and the initial helium abundance $Y$, integrated in the luminosity range $-1.6 \leq M_{I} \leq 1.4$.  In the top rows we list the model outputs for scaled-solar abundances and ages $t=12$ Gyr.  In the middle rows we show the model outputs for $\alpha$-enhanced isochrones, which are marginally different than scaled-solar isochrones at fixed [M/H], age, and initial helium abundance. In the bottom rows we list the model outputs for a range of ages and helium abundances, due to the uncertainty in the age-helium-metallicity relation of bulge stars at high metallicity.    \newline}
\centering 
\begin{tabular}{ccccccccccccccc}
	\hline \hline
[M/H] & [$\alpha$/Fe] & $Y$ & t/Gyr & $B$ & $EW_{RC}$ & $M_{I,RC}$ &   $(V-I)_{RC}$ &   $(V-K)_{RC}$  & $(M/M_{\odot})_{RC}$ & ${\log{g}}_{RC}$   &     $f^{RC}_{RGBB}$ & ${\Delta}I^{RC}_{I_{RGBB}}$  & $f^{RC}_{AGBB}$ & ${\Delta}I^{RC}_{I_{AGBB}}$ \\
	\hline \hline \hline
$-$1.27  & 0.0  & 0.25 &  12 &  0.66 & 1.75 & $-$0.29  & 0.78 &    1.75 &      0.74 & 2.46  & 0.10 &  $-$0.47 & 0.02 &  $-$1.03 \\ \hline
$-$0.66  & 0.0     & 0.25 &  12 &  0.67 & 1.98 & $-$0.31 &   0.90 & 2.05 &        0.78 &  2.37 &      0.18 &   0.17 & 0.04 &  $-$1.04 \\ \hline
$-$0.35  & 0.0     & 0.26 &  12 &  0.65 & 2.08 & $-$0.27 & 0.98 &  2.23 &         0.82 &    2.34 &             0.23 &  0.44 & 0.03 &  $-$1.04 \\ \hline
$-$0.25  & 0.0     & 0.26 &  12 &  0.65 & 2.12 & $-$0.24 &  0.99 &   2.29 &       0.84 &    2.35 &       0.25 &  0.51 & 0.04 &  $-$1.04 \\ \hline
$+$0.06  & 0.0     & 0.27 &  12 &  0.61 & 2.10 & $-$0.17 &  1.08 & 2.49 & 0.91 & 2.33 &    0.29 &  0.66 & 0.04 &  $-$1.04 \\ \hline
$+$0.25  & 0.0     & 0.29 &  12 &  0.62 & 2.26 & $-$0.13 &  1.15 & 2.61 & 0.94 &  2.32 &         0.30 &  0.78 & 0.05 &  $-$1.02 \\ \hline
$+$0.40  & 0.0     & 0.30 &  12 &  0.58 & 2.32 & $-$0.12  & 1.20 & 2.71 &   0.95 &  2.30  &  0.30 &   0.85 & 0.05 &  $-$0.98 \\ \hline	\hline
$-$1.27  & +0.4     & 0.25 &  12 &  0.65 & 1.75 & $-$0.29 &  0.76 & 1.72 & 0.74 & 2.47 &    0.08 &  $-$0.53 & 0.02 &  $-$1.03 \\ \hline
$-$0.66  & +0.4     & 0.25 &  12 &  0.66 & 2.02 & $-$0.33 &  0.88 &   2.00 &       0.77 &    2.37 &       0.16 &  0.11 & 0.03 &  $-$1.02 \\ \hline
$-$0.35  & +0.4     & 0.26 &  12 &  0.66 & 2.18 & $-$0.31 &  0.95 & 2.17 & 0.80 &  2.34 &         0.21 &  0.41 & 0.02 &  $-$1.03 \\ \hline
$+$0.06  & +0.4     & 0.27 &  12 &  0.60 & 2.21 & $-$0.25  & 1.07 & 2.31 &   0.87 &  2.30  &  0.26 &   0.66 & 0.02 &  $-$1.02 \\ \hline	\hline
$+$0.26  & 0.0     & 0.35 &  11 &  0.60 & 2.86 & $-$0.31 &  1.08 &    2.43 &      0.72 &  2.20  &     0.20 &   0.80  & 0.05 &  $-$1.12 \\ \hline
$+$0.26  & 0.0     & 0.32 &  11 &  0.59 & 2.50 & $-$0.17  & 1.11 &   2.50 &       0.78 & 2.27     & 0.25 &   0.69 & 0.05 &  $-$1.12\\ \hline
$+$0.26  & 0.0     & 0.32 &  7 &  0.58 & 2.64 & $-$0.28  &  1.12 &    2.58 &       0.95 & 2.30 &       0.19 &   0.59 & 0.04 &  $-$1.05 \\ \hline
$+$0.25  & 0.0     & 0.29 &  7 &  0.59  & 2.32 & $-$0.21  & 1.14 &  2.56 &        1.12 &   2.39          & 0.22 &   0.57 & 0.04 &  $-$0.99 \\ \hline
$+$0.25   & 0.0    & 0.29 &  4 &  0.61 & 2.33 & $-$0.27 & 1.12 &      2.53 &       1.32 & 2.46 & 0.15 &    0.44 & 0.06 &  $-$0.97 \\ \hline
	\hline
\end{tabular}
\label{table:PredictedLuminosityParameters}
\end{table*}

\section{Models}
\label{sec:Models}
The stellar isochrones and luminosity functions used in this work are predominantly taken from the BaSTI stellar database\footnote{\label{foot:1}http://albione.oa-teramo.inaf.it}. The models assume a scaled-solar abundance mixture without overshooting, and include both the first \citep{2004ApJ...612..168P} and second \citep{2007AJ....133..468C} ascent  of the RG branch, hereafter respectively referred to as the RGB and asymptotic giant branch (AGB). Additional models with enhanced helium-enrichment ($Y=0.32,0.35$) that assume otherwise identical physics to those downloaded from BaSTI database have been computed specifically for this work. The predicted parameters of the combined RG, RC, and AGB luminosity function  are listed in Table \ref{table:PredictedLuminosityParameters}.

The N-body model used in this work has been made by \citet{2003MNRAS.341.1179A} and used by \citet{2012ApJ...756...22N}, where the latter estimated a scale factor between model distance units and Kpc of 1.2 as optimal to interpret Galactic kinematics data. The model is initialised as an isolated, axisymmetric galaxy with a live disk and halo component, with the disk stars having an exponential distribution in the radial direction. The bar grows from the disk and part of it buckles to form an X-shaped structure (for a review of the process, see \citealt{2005MNRAS.358.1477A}). We evaluate the model in the same evolutionary state as \citet{2012ApJ...756...22N}. We assume a distance between the Sun and the Galactic centre of 8.13 Kpc, and a viewing angle between the major axis of the Galactic bar and the line of sight between the Sun and the Galactic Centre of $\alpha=29.4^{\circ}$ \citep{2013MNRAS.434..595C}. 

\section{First Systematic Bias: The Horizontal Branch Becomes Bluer with Decreasing Metallicity, Eventually Being Selected Against}
\label{sec:Systematic1}
\begin{figure}
\begin{center}
\includegraphics[totalheight=0.35\textheight]{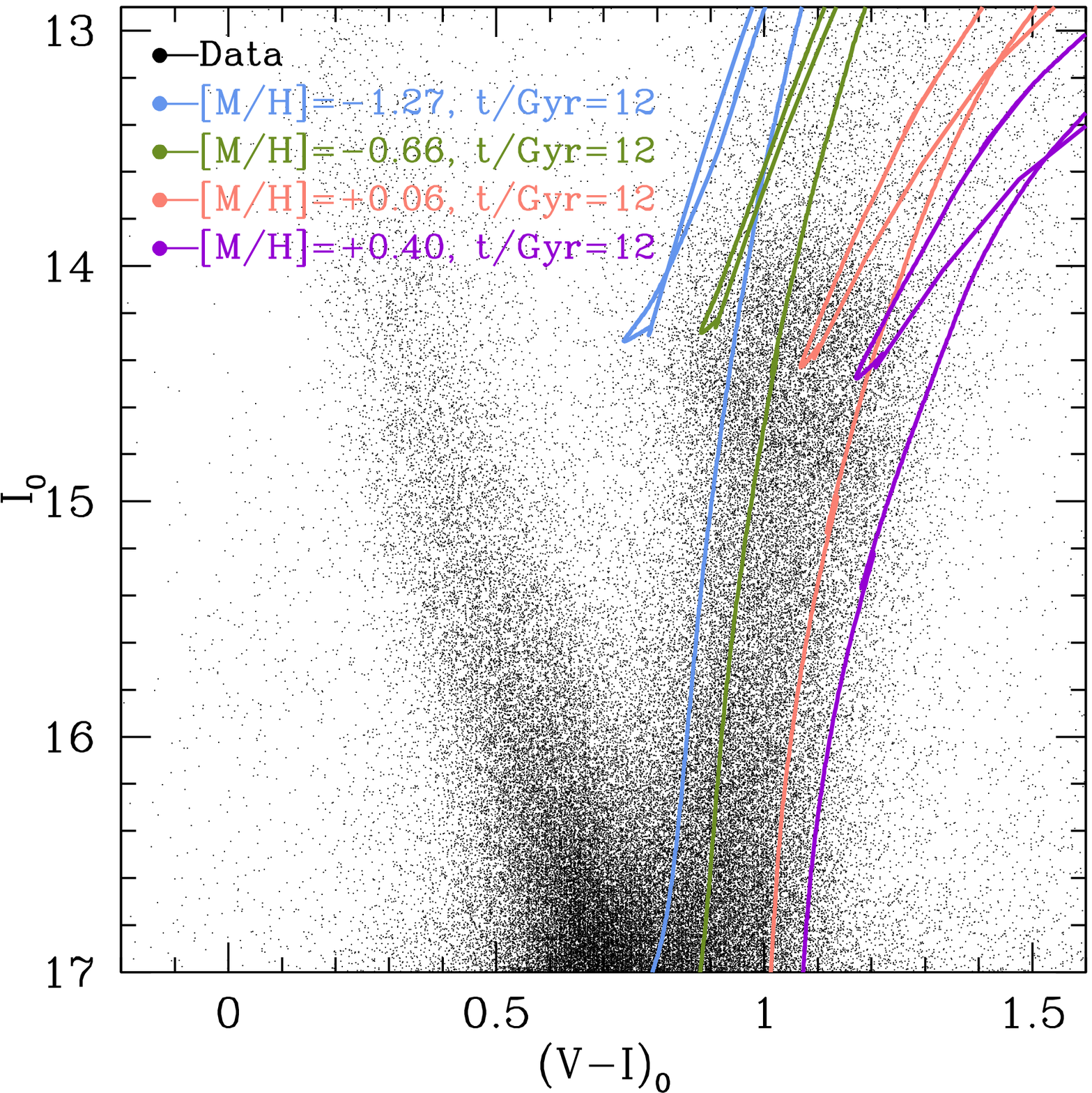}
\end{center}
\Large
\caption{\large BaSTI\textsuperscript{\ref{foot:1}}  t$=12$ Gyr isochrones of metallicities [M/H]$=-1.27,-0.66,+0.06,+0.40$ \citep{2004ApJ...612..168P,2007AJ....133..468C} over plotted on a dereddened $(V-I,I)$ Galactic bulge colour-magnitude diagram toward the OGLE-III field BLG16 \citep{2011AcA....61...83S,2013ApJ...769...88N}, centred on $(l,b)=(0.00^{\circ},-5.80^{\circ})$. We assume a distance modulus of $\mu=14.55$ for the overplotting of the isochrones. The metallicity distribution function of any bulge sample as well as the ratio of red clump stars to red giant stars will clearly be a sensitive function of the colour selection. } 
\label{Fig:BastiIsochroneFigure}
\end{figure}

The first systematic bias that we quantify arises from the fact that the RGB is redder than the RC at fixed metallicity and luminosity ($(V-I)_{RG} - (V-I)_{RC} \approx 0.15$ mag) and that both become bluer with decreasing metallicity, at a rate of $d(V-I)_{RC}/d\rm{[M/H]} \approx 0.25$ mag dex$^{-1}$ -- see Figure \ref{Fig:BastiIsochroneFigure} and Table \ref{table:PredictedLuminosityParameters}. This means that the ratio of RC stars to RG stars will be a metallicity-dependent function of the colour-cut used by any given survey, modifying the diagnostic power of a sample to resolve the distance distribution function. That is because unlike RC stars, RG stars have an intrinsic luminosity function dispersed over $\sim$4 magnitudes, and are thus much less suitable to distance determinations. 

Reviewing the potential impact of this effect on current results: \begin{itemize}
\item  \citet{2012ApJ...756...22N} use a colour-cut of $(J-K)_{0} \geq 0.40$ (corresponding to $(V-I)_{0} \gtrsim$ 0.65), and as such should  have neutrally-sampled the bulk  ($\gtrsim 99$\%) of bulge RC and RG stars, given the assumption that their stellar evolution is adequately predicted by the models used in this work;
\item \citet{2012A&A...546A..57U} use a selection that imposes an effective colour-cut at the RC of $0.60  \lesssim (J-K)_{0}  \lesssim 0.70$, corresponding to $0.97  \lesssim (V-I)_{0}  \lesssim 1.13$. This will have the effect of decreasing the ratio of RC to RG stars at the metal-poor end, and increasing it at the metal-rich end. \citet{2012A&A...546A..57U} indeed reports more prominent RCs for [M/H] $\geq -0.20$ than for [M/H] $< -0.20$;
\end{itemize}

We note that the model predictions stated in Table \ref{table:PredictedLuminosityParameters} may be overestimating the colour of metal-poor horizontal branch stars in the bulge. The Galactic bulge RR Lyrae (ab type) population has a mean metallicity of [Fe/H]$\approx -1.00$ \citep{2012ApJ...750..169P}, which indicates that metal-poor horizontal branch stars in the bulge may be bluer than predicted by a whopping ${\delta}(V-I) \approx 0.35$ mag. Galactic bulge globular clusters are also known to have horizontal branches that are very blue for their metallicities \citep{2009A&A...507..405B}. \citet{1992AJ....104.1780L} argues that the colour of huge metal-poor horizontal branch stars indicates an old bulge, though we find that an unphysical age of $t \approx 17.0$ Gyr is needed to produce an RRab morphology for stars with [Fe/H]$=-1.0$ if one assumes standard stellar physics. A plausible explanation is that the mass-loss is greater-than-expected for low-metallicity RG stars in the bulge, leading to a bluer horizontal branch morphology. Alternatively, the helium-abundance of these stars might be higher than expected. Until this discrepancy has been explained, it will not be possible to quantify selection biases for low-metallicity ([Fe/H] $\lesssim -0.80$)  bulge horizontal branch stars. 

In addition, the referee directs us to a curious discrepancy shown in Figure 20 of \citet{2003A&A...399..931Z}, where synthetic modelling of the bulge luminosity function predicts a mean RC colour that is $\sim$0.1 mag bluer in $(J-K)$ than the RGB at the same apparent magnitude in $K$, whereas there no such offset in the observed CMD. This discrepancy suggests that it will be difficult to model the selection effects. We agree that this is a cause for concern that would benefit from further investigation. We argue that it is likely due to errors at higher metallicity ([Fe/H] $\gtrsim$ 0), where each of stellar mass-loss along the red giant branch, colour-temperature relations, and spectroscopic model atmospheres are more theoretically uncertain, as well as the fact that the age and helium abundance of the bulge are empirically uncertain at higher metallicities. We think it is unlikely that this is an issue at lower metallicities. The same theoretical framework adopted in present investigation appears fully appropriate to describe the RC morphology of stars in the solar neighbourhood as investigated with the Hipparcos satellite \citep{2004ApJ...612..168P} and the horizontal branch morphology of metal-poor Galactic globular clusters (see, for instance, \citealt{2013MNRAS.430..459D}, and references therein). Separately, Figure 20 of \citet{2003A&A...399..931Z} also shows an excess in observations relative to predictions in the number of horizontal branch stars bluer than the RC, in agreement with  \citet{1992AJ....104.1780L}.


\section{Second Systematic Bias: The Ratio of Red Clump to Red Giant Stars is an Increasing Function of Metallicity}
\label{sec:Systematic2}
Higher metallicity has the effect of decreasing the duration of the RGB, and of increasing the lifetime of the core helium-burning phase (see discussions in \citealt{1994A&A...285L...5R}, \citealt{2000ApJ...538..289Z}, and \citealt{2006essp.book.....S}). This means that a metal-rich sample will have a higher ratio of RC to RG+AGB stars, and thus a more easily discernible distance distribution function.  We parameterise this effect by measuring the equivalent width of the RC on the luminosity function in the models, $EW_{RC}$, which is the ratio of the number of RC stars to the combined number density of RG and AGB stars at the luminosity of the RC. It is similar to the parameter $EW_{RGBB}$ previously measured for Galactic globular clusters by \citet{2013ApJ...766...77N}, though for bulge stars one cannot separate the AGB from the RGB, thus leading to a different normalisation -- which also effects determinations of the parameter $B$. 

We find that for a scaled-solar helium abundance and fixed age ($t=12$  Gyr),  the equivalent width of the RC rises from $EW_{RC} = 1.75$ at [M/H]$= -$1.27 to $EW_{RC} = 2.08$ at [M/H]$= -0.35$, and finally to $EW_{RC} = 2.32$ at [M/H]$= +$0.40.  Thus, the ratio of RC to RGB+AGB stars rises by $\sim$33\% as the metallicity increases from [M/H]$=-1.27$ to [M/H]$=+0.40$. In other words, in the limiting case of metallicity and kinematics being uncorrelated, a sample of bulge stars with [M/H]$=-1.27$ at the apparent luminosity of the RC would need to be $1.33^{2} = 1.76 {\times}$ larger than the corresponding sample of stars with [M/H]$=+0.40$ (since statistical significance typically scales as the square root of the number of data points), in order to measure features such as the bimodality in the distance distribution function with comparable statistical significance, all other factors being equal (which they're not, see Section \ref{sec:Systematic3}). We note that the analysis of \citet{2012ApJ...756...22N}  should not be affected by this bias, as their population partition  included 80 stars for [Fe/H[$\geq 0$, 240 stars for $-0.50 \leq$[Fe/H[$<0$, and 200 stars for [Fe/H[$< -0.50$ -- \citet{2012ApJ...756...22N}   already have more stars in their metal-poor bins, due to where they set the bincenters and the fact they studied sightlines relatively far from the plane, where the mean metallicity is lower. 

The bias estimated by this section is in fact a lower bound, as reports in the literature argue that metal-rich bulge stars may be, on average, younger \citep{2013A&A...549A.147B} or helium-enhanced \citep{2012ApJ...751L..39N,2013ApJ...766...77N}, or both younger and helium-enhanced \citep{2013MNRAS.428.2577B}, all of which would further increase $EW_{RC}$ at high metallicity.  Predicted parameters for these scenarios are listed in the lower part of Table \ref{table:PredictedLuminosityParameters}.

\section{Third Systematic Bias: The Prominence of the Red Giant Branch Bump is an Increasing Function of Metallicity}
\label{sec:Systematic3}
The RC is not the only departure from an exponential continuum in the luminosity function of RG stars. During the RGB, there is a relatively brief period where the nuclear efficiency of the hydrogen burning shell temporarily drops, leading to a corresponding decrease in the luminosity of the star, thus leading to an excess in the luminosity function called the ``red giant branch bump" (RGBB, as before). The number counts and characteristic luminosity of the RGBB are a steeply sensitive function of the age, metallicity, and helium abundance of a stellar population \citep{1997MNRAS.285..593C,2010ApJ...712..527D,2013ApJ...766...77N}. We describe the metallicity-dependence of the predicted parameters for the RGBB in this section, and then we demonstrate that failure to account for this component of the luminosity function will bias determinations of the distance distribution function. 

Fixing the age to $t=12$ Gyr and the helium abundance to scaled-solar, the predicted luminosity of the RGBB relative to the RC, ${\Delta}I^{RC}_{I_{RGBB}}$, decreases from ${\Delta}I^{RC}_{I_{RGBB}}=-0.47$ to ${\Delta}I^{RC}_{I_{RGBB}}=+0.85$ as the metallicity increases from [M/H]$=-1.27$ to [M/H]$=+0.40$, an impressive shift of 1.32 mag in luminosity over 1.67 dex in metallicity. In contrast, the separation in brightness between the two RCs of the Galactic bulge is $\sim$0.45 mag \citep{2010ApJ...721L..28N,2010ApJ...724.1491M,2013ApJ...776...76P}, approximately corresponding to ${\Delta}I^{RC}_{I_{RGBB}}$ for stars with metallicity [M/H]$=-0.35$. In addition to the decreasing luminosity of the RGBB, the ratio of RGBB to RC stars, $f^{RC}_{RGBB}$, is predicted to increase from $f^{RC}_{RGBB}=0.10$ to $f^{RC}_{RGBB}=0.30$ over the same metallicity range. It is therefore straightforward to understand why ignoring the RGBB will thus lead one to infer distorted brightness differences between the two RCs, with the effect growing worse with increasing metallicity. The predicted parameters of the RGBB are listed in Table \ref{table:PredictedLuminosityParameters}, where we also list the corresponding parameters for the asymptotic giant branch bump (AGBB).  

\begin{figure*}
\begin{center}
\includegraphics[totalheight=0.65\textheight]{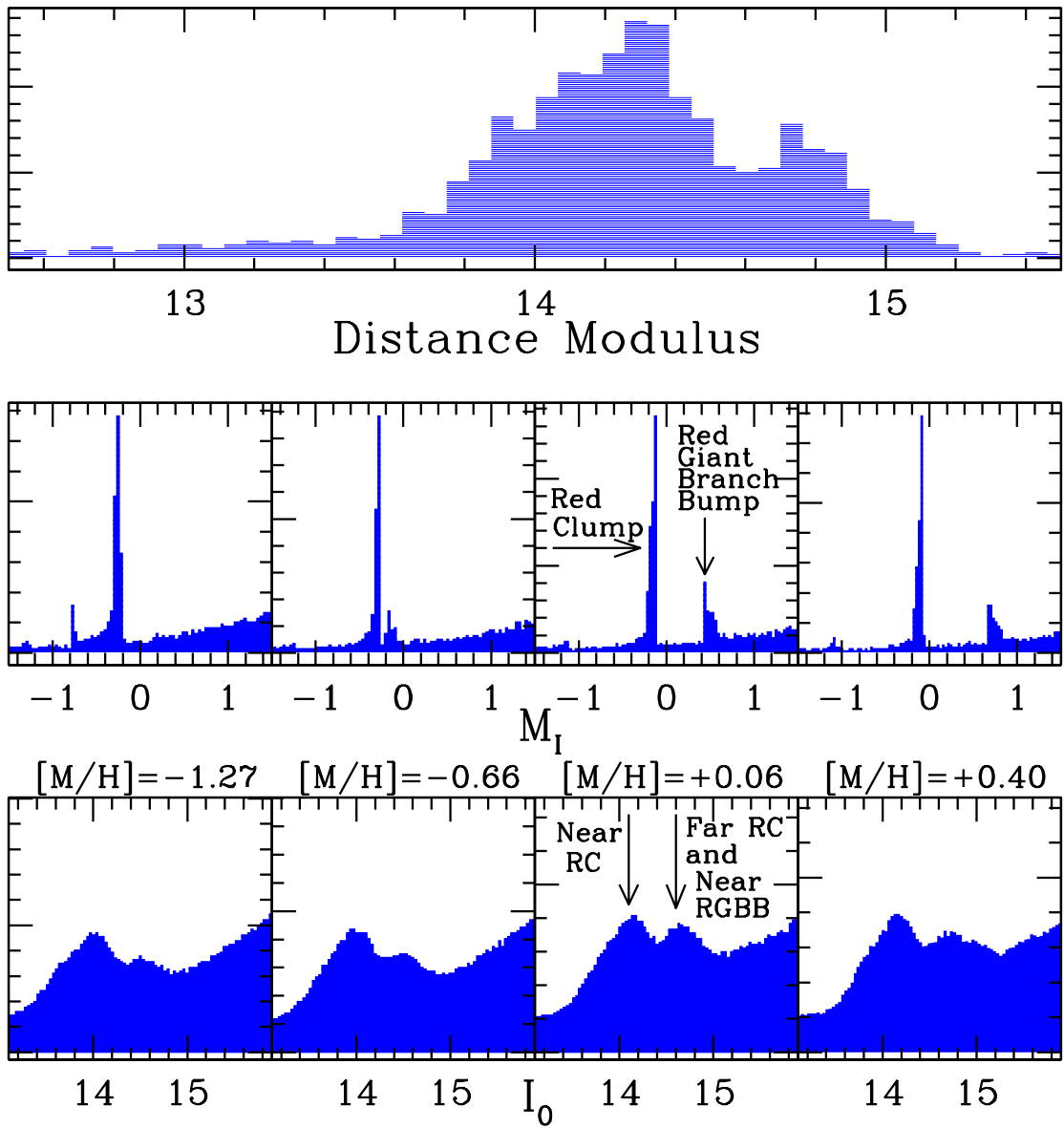}
\end{center}
\Large
\caption{\large TOP: Predicted distance distribution toward $(l,b)=(0.00^{\circ},-10.00^{\circ})$ from the N-body model \citep{2003MNRAS.341.1179A}, a sightline with an unambiguous bimodality in its distance distribution function. MIDDLE:  BaSTI  $t=12$ Gyr luminosity functions for the red giant branch as a function of [M/H] \citep{2004ApJ...612..168P,2007AJ....133..468C}. BOTTOM: Convolution of the distance distribution and absolute magnitude distribution function to produce apparent magnitude distribution functions for the four metallicities. Without accounting for the metallicity-dependence of the stellar luminosity function, the red clump will appear more split at higher metallicities due to the similar apparent magnitudes of the near red giant branch bump and the far red clump. } 
\label{Fig:NBodyLF}
\end{figure*}

In Figure \ref{Fig:NBodyLF}, we convolve the distance modulus distribution for $(l,b) = (0^{\circ},-10^{\circ})$ (a sightline used by both \citealt{2012ApJ...756...22N}   and  \citealt{2012A&A...546A..57U}) predicted by our N-body model (top panel) with intrinsic luminosity functions corresponding to four different metallicities ([M/H]$=-1.27,-0.66+0.06,+0.40$, middle panels), and Gaussian noise of $\sigma_{I}=0.07$ mag to simulate the effects of photometric errors and differential extinction to produce four predicted apparent luminosity functions (bottom panels).  We fit the final luminosity functions using the same methodology as \citet{2013ApJ...776...76P}. If we ignore the RGBB in our fits, fitting for a split red clump leads to ${\chi}^2$ reductions of \{14\%, 33\%, 61\%, 74\%\}  in the four metallicity bins relative to the ${\chi}^2$ value obtained when fitting a single RC. In other words, there is the illusion that the distance modulus distribution is more cleanly bimodal among the metal-rich stars, when in fact the increased signal partly emerges from differences in stellar evolution in this construction. This is also rather obvious simply by inspecting the bottom-four panels of Figure \ref{Fig:NBodyLF}. More simply, the RGBB of the nearer (brighter) component has a similar brightness to the RC of the further (fainter) component, which creates a misleading amplification the signal-to-noise of the fainter peak. 

\subsection{The Effect of a Surface Gravity Cut on Red Giant Branch Bump Contamination}
\label{sec:Systematic3logg}
In order to have a purer sample of RC stars, \citet{2012ApJ...756...22N} limited their analysis to stars with $1.90 \leq \log{g} \leq 3.10$. We simulate the effect of this selection by constructing a luminosity function whereby for each star we ``measure" $\log{g}$ with a Gaussian error of 0.3 dex, keeping only those stars in the interval $2.00 \leq \log{g} \leq 3.00$. We find that this will not impact the degree of RGBB contamination at high metallicities. For the $t=12$ Gyr, [M/H]$=+0.06$ isochrone, the value of $f^{RC}_{RGBB}$ increases from 29\% to 30\%, whereas for the $t=12$ Gyr, [M/H]$=+0.25$ isochrone $f^{RC}_{RGBB}$ decreases from 29\% to 27\% -- both small changes, reflective of the fact that $\log{g}_{RGBB}$ is typically $\sim 2.60$ at high metallicities, and thus within the measurement error of $\log{g}_{RC}$. 

The observational fact that this selection improved the clarity of the sample (see Figure 3 of \citealt{2012ApJ...756...22N}) is more likely due to it decreasing the rate of disk contamination. As disk stars are on average closer than the bulge, their contribution to the luminosity function at the apparent magnitude of the RC will be from stars dimmer than the RC, which are more numerous.

\section{Discussion and Conclusion}
\label{sec:Discussion}
In this work, we have demonstrated that deriving accurate cartography of the X-shape and its possible metallicity-dependence necessitates a rigorous treatment of not only the spatial morphology, but of stellar physics as well. We explored in detail the predicted metallicity-dependence of the colours of the RC and RGB, of the ratio of RC to RG stars, and of the RGBB and their effects on studies of the split RC. 
 
This requirement to rigorously treat stellar evolution has previously been acknowledged by at least some of the literature. Both \citet{2010ApJ...721L..28N} and \citet{2010ApJ...724.1491M}, in their independent discovery papers of the split RC, noted that the RGBB would confuse measurements of the properties of the two RCs. \citet{2013ApJ...776...76P} and  \citet{2013MNRAS.435.1874W} both included the RGBB as part of their parameterization in modelling the spatial distribution function of RC stars. \citet{2013A&A...555A..91V} used BaSTI \citep{2004ApJ...612..168P,2007AJ....133..468C} isochrones to estimate an 11\% cross-contamination rate between their bright and faint RC spectroscopic samples.  In contrast, \citet{2011AJ....142...76S}  ignored the RGBB in their parameterization. We expect that as further analysis is completed, the values of the brightness difference between the two peaks and the fraction of stars in the fainter RC suggested by \citet{2011AJ....142...76S} will be shown to be overestimated. 

We note that even with the precise predictions listed in Table \ref{table:PredictedLuminosityParameters} it will not be straightforward to convolve N-body models with stellar models to simulate apparent magnitude distribution functions. The first issue is that the luminosity of the RGBB has been shown to be likely overestimated by conventional stellar models by $\sim$0.20 mag with a possible metallicity-trend in the offset \citep{2010ApJ...712..527D,2011PASP..123..879T,2011A&A...527A..59C}. The second issue is that even if stellar models were perfect, it would still not be clear \textit{which} stellar models to actually use, as the age-helium-metallicity relation of the bulge is uncertain at high metallicities  \citep{2013A&A...549A.147B,2012ApJ...751L..39N,2013ApJ...766...77N,2013MNRAS.428.2577B}.

Finally, we comment on two sources of uncertainty not explored in this work. The first is that of contamination from foreground or background disk stars not in bar orbits but with apparent magnitude distributions overlapping those of stars in the bar. There will be disk contamination at the luminosity of the RC within any bulge photometric sample, and further, that ratio could be metallicity-dependent. As the disk stars have a different spatial distribution than stars captured around bar/bulge orbits, this will lead to distortions in the distance distribution function.  The second source of uncertainty lies with the shape of the distance distribution function. Each of \citet{2010ApJ...721L..28N}, \citet{2012A&A...546A..57U}, \citet{2013ApJ...776...76P} and  \citet{2013MNRAS.435.1874W} investigated the split RC by assuming a Gaussian distribution for the apparent magnitudes of the two RCs.  However, the top panel of Figure \ref{Fig:NBodyLF} predicts non-Gaussian distance distribution functions along the line of sight, with a negative skew for the brighter RC and a positive skew for the fainter RC. For a skewed distribution function, the mode will not correspond to the mean, which could lead to distortions when comparing data to models, or when simply fitting for the RCs in the luminosity function. 

The mapping of the spatial morphology of Galactic bulge stars and the extent to which the mapping depends on metallicity is a fundamental research enterprise in Galactic archeology. However, this enterprise is a challenging one, with numerous systematics potentially plaguing the way forward. As more data (photometry, spectroscopy, proper motions, etc) comes in from surveys such as Gaia-ESO \citep{2012Msngr.147...25G}, GIBS \citep{2014arXiv1401.4878Z}, VVV\citep{2012A&A...537A.107S} and OGLE-IV \citep{2012AcA....62..219S}, we expect not only better diagnostic power, but also the need for more sophisticated accounting of stellar physics and Galactic dynamics to properly interpret these data.


\section*{Acknowledgments}
We thank the referee, Manuela Zoccali, for a helpful report and comments. We thank M. Ness, S. Uttenthaler, and M. Asplund for helpful discussions. 

DMN was  supported by the Australian Research Council grant FL110100012. SC is grateful for financial support from PRIN-INAF 2011 "Multiple Populations in Globular Clusters: their role in the Galaxy assembly" (PI: E. Carretta), and from PRIN MIUR 2010-2011, project \lq{The Chemical and Dynamical Evolution of the Milky Way and Local Group Galaxies}\rq, prot. 2010LY5N2T (PI: F. Matteucci).   EA acknowledges financial support from the CNES (Centre National d'Etudes Spatiales - France) and from the People Programme  (Marie Curie Actions) of the European Union's Seventh Framework Programme FP7/2007-2013/ under REA grant agreement number PITN-GA-2011-289313 to the DAGAL network.   EA is thankful for HPC resources from GENCI- TGCC/CINES (Grants 2013 - x2013047098 and 2014 - x2014047098).

This work has made use of BaSTI web tools. 


\end{document}